\documentclass[epj,referee]{svjour}
\usepackage{amssymb}
\usepackage{latexsym}
\usepackage{graphics}
\begin{document}
\title{
Universal scaled Higgs-mass gap
for the bilayer Heisenberg model
in the ordered phase
}
\author{Yoshihiro Nishiyama  
}                     
\offprints{}          
\institute{Department of Physics, Faculty of Science,
Okayama University, Okayama 700-8530, Japan}
\date{Received: date / Revised version: date}
%
\abstract{
The
spectral properties for the bilayer quantum Heisenberg model
were investigated with
the numerical diagonalization
method.
In the ordered phase, 
there appears 
the massive
Higgs excitation embedded in the continuum of the Goldstone excitations.
Recently,
it was claimed that the 
{\it properly scaled}
Higgs mass is a universal constant in proximity to the critical point.
Diagonalizing
the finite-size cluster with $N \le 36$ spins,
we calculated the dynamical scalar susceptibility $\chi_s''(\omega)$, which is 
rather
insensitive to the Goldstone mode.
A finite-size-scaling analysis of $\chi_s ''(\omega)$ 
is made,
and the universal (properly scaled)
Higgs mass is estimated.
\PACS{
{75.10.Jm}        {Quantized spin models} \and 
{05.70.Jk} {Critical point phenomena} \and
{75.40.Mg} {Numerical simulation studies} \and
{05.50.+q} {Lattice theory and statistics (Ising, Potts, etc.)}
     } 
} 
\maketitle

\section{\label{section1}Introduction}

In
the spontaneous-symmetry-breaking phase,
there appear
the Goldstone and Higgs excitations
in the low-energy spectrum.
The former (latter) excitation is massless (massive),
and the continuum of the former overwhelms
the latter dispersion branch.
[In regard to the wine-bottle bottom potential,
the former (latter) excitation corresponds to the
azimuthal (radial) modulation of the order parameter concerned.]
Recently,
the Higgs-excitation spectral peak
was observed for the
the
two-dimensional ultra-cold atom 
\cite{Endres12}.
Here, a key ingredient is 
that the external disturbance,
namely, the trap-potential modulation,
retains the U$(1)$ (gauge) symmetry,
and it is rather insensitive to the 
(low-lying) Goldstone modes.
Moreover, the experiment revealed 
a gradual closure of the Higgs mass $m_H$
in proximity to the
critical point
between
the superfluid and Mott-insulator phases;
the criticality 
belongs to the $(2+1)$-dimensional O$(2)$ universality class,
and the singularity lies out of the scope of
the Ginzburg-Landau theory.
Such O$(2)$- [equivalently, U$(1)$-] symmetric system is
ubiquitous in nature,
and the underlying physics is common to a wide variety of substances;
we refer readers to Ref. 
\cite{Pekker15} for a review.

In this paper, we investigate the O$(3)$-symmetric counterpart,
namely, 
the bilayer quantum Heisenberg model 
\cite{Matsushita97,Troyer98,Sommer01}, 
by means of the numerical diagonalization method.
Our aim is to estimate the scaled Higgs mass
(universal amplitude ratio)
$m_H/\Delta$ ($\Delta$: the excitation gap in the adjacent paramagnetic phase);
technical details are addressed in Sec. \ref{section2}.
The scaled Higgs mass has been
estimated as $m_H/\Delta=2.2(3)$ \cite{Gazit13a} and 
$2.6(4)$ \cite{Lohofer15} with the (quantum) Monte Carlo method.
On the one hand, via the elaborated renormalization-group analyses,
the scaled Higgs mass was estimated as
$m_H/\Delta=2.7$ \cite{Rose15}
and 
$1.64$ \cite{Katan15}.
An advantage of the numerical diagonalization approach
is that the spectral property is accessible
directly
\cite{Gagliano87} without resorting to the inverse Laplace transformation
(see Appendix B of Ref. \cite{Gazit13b}).
It has to be mentioned that the scaled Higgs mass
$m_H / \Delta$
has been investigated extensively
as for the O$(2)$-symmetric case,
\cite{Gazit13a,Rose15,Katan15,Gazit13b,Chen13,Rancon14,Nishiyama15}.
According to the study
\cite{Rancon14}
of the O$(N)$-symmetric system with generic $N$,
the Higgs-excitation peak should get broadened 
for large $N$.

\begin{figure}
\includegraphics
{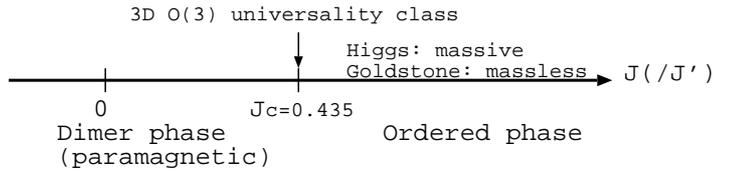}%
\caption{\label{figure1}
A schematic phase diagram \cite{Matsushita97,Troyer98}
for the bilayer Heisenberg model 
(\ref{Hamiltonian})
is presented.
A critical point locates at $J_c=0.435$ \cite{Troyer98},
separating the paramagnetic and ordered phases.
In the latter phase,
the Higgs (Goldstone) excitation is massive (massless).
The Higgs-excitation peak for the spectral function
(\ref{scalar_susceptibility})
is our concern.
}
\end{figure}

To be specific,
we present the Hamiltonian for the bilayer Heisenberg model
\cite{Matsushita97,Troyer98,Sommer01}
\begin{equation}
\label{Hamiltonian}
{\cal H}=
-J \sum_{a=1}^2 \sum_{\langle ij\rangle} {\bf S}_{ai}\cdot{\bf S}_{aj}
+J'\sum_{i=1}^{N/2} {\bf S}_{1i} \cdot {\bf S}_{2i}
   .
\end{equation}
Here, the spin-$S=1/2$ operator 
${\bf S}_{ai}$ is placed at each square-lattice point 
$i$ ($i=1,2,\dots,N/2$)
within each layer $a$ ($a=1,2$)
The summation $\sum_{\langle ij\rangle}$
runs over all possible nearest neighbor pairs 
$\langle ij \rangle$
within each layer.
The parameter $J(>0)$ [$J'(>0)$] denotes the
intra- (inter-) layer ferromagnetic (antiferromagnetic) Heisenberg
interaction;
hereafter, we consider $J'$ as the unit of energy ($J'=1$).
The phase diagram \cite{Matsushita97}
is presented in Fig. \ref{figure1}.
At $J_c=0.435$ \cite{Troyer98}, there occurs a phase transition,
separating the paramagnetic ($J<J_c$) and 
ordered ($J>J_c$) phases;
the phase transition
belongs to the three-dimensional O$(3)$ universality class \cite{Troyer98}.
The criticality of the spectral function in the ordered phase
is our concern.
It has to be mentioned that the 
recent quantum Monte Carlo simulation 
\cite{Lohofer15}
also
treats the bilayer Heisenberg model (\ref{Hamiltonian}),
albeit with an antiferromagnetic intra-layer interaction,
$J<0$.
The setting of the interaction parameter may be arranged 
suitably for each methodology.
Nevertheless,
details of magnetism should
not influence 
the criticality
of $m_H$.

The rest of this paper is organized as follows.
In Sec. \ref{section2},
we analyze the spectral properties
for the Hamiltonian
(\ref{Hamiltonian})
by means of the numerical diagonalization method.
The simulation algorithm is presented in Appendix.
In Sec. \ref{section3},
we address the summary and discussions.

\section{\label{section2}
Numerical results
}

In this section,
we present the numerical results.
We employed the numerical diagonalization method
for the finite-size cluster with
$N \le 36$ spins.
We implemented the screw-boundary condition 
(Appendix) \cite{Novotny90} in order to treat a variety of $N=30,32,\dots$
systematically.
Because the $N$ spins constitute the bilayer cluster,
the linear dimension of the cluster is given by
\begin{equation}
L=\sqrt{N/2} .
\end{equation}

\subsection{\label{section2_1}
Finite-size-scaling analysis of the Goldstone mass $m_G$}

As a preliminary survey,
we analyze the 
Goldstone mass $m_G$ with the finite-size-scaling method.
The Goldstone mass $m_G$ is identified as the triplet (magnon) excitation gap;
hence, the simulation was performed within the
subspace specified by the longitudinal
total magnetization,
either $S^z_{tot}=0$ or $1$.

\begin{figure}
\includegraphics{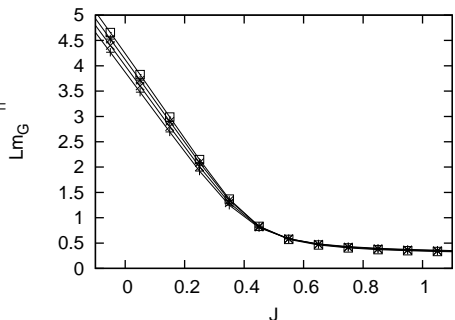}%
\caption{\label{figure2}
The scaled Goldstone mass $L m_G$ is plotted for
various $J$ and 
($+$) $N(=2 L^2)=30$,
($\times$) $32$,
($*$) $34$ and
($\Box$) $36$.
The result agrees with the preceding estimate
$J_c=0.435$
\cite{Troyer98};
note that
the intersection point of the curves indicates
the location of the critical point.
In the ordered phase $J>J_c$,
the Goldstone-excitation gap closes, 
$m_G [< {\rm O}(L^{-1})]\to 0$,
in the thermodynamic limit,
$L \to \infty$.
On the contrary,
in the paramagnetic phase $J<J_c$,
an excitation gap opens;
the gap [see Eq. (\ref{Delta})] sets a fundamental energy scale 
for the subsequent scaling analyses.
}
\end{figure}

In Fig. \ref{figure2},
we present the scaled Goldstone mass $Lm_G$
for various $J$ and system sizes, 
$N(=2 L^2)=30$ ($+$),
$32$ ($\times$)
$34$ ($*$)
and
$36$ ($\Box$).
The data suggest that a phase transition
takes place at $J\approx0.4$;
note that the intersection point of the curves
indicates the location of the critical point.
In Ref. \cite{Troyer98},
the critical point was estimated as
$J_c=0.435$; our result agrees with this estimate.
The Goldstone mass appears to close,
$m_G[<{\rm O} (L^{-1})]\to 0$,
in the ordered phase $J>J_c$ as $ L \to\infty$
(thermodynamic limit).
On the contrary, in the paramagnetic phase $J<J_c$,
a finite mass gap opens;
in the quantum-magnetism language,
the mass gap is interpreted as 
the spin (magnon) gap for the spin-liquid phase.
The spin gap [see Eq. (\ref{Delta})]
sets a fundamental energy scale
for the subsequent scaling analyses.

\begin{figure}
\includegraphics{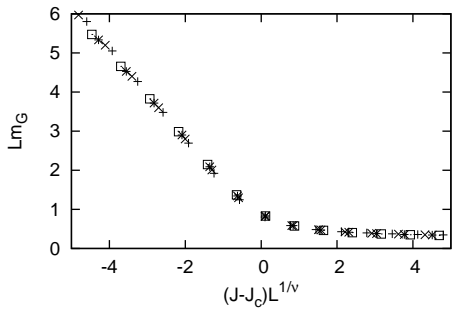}%
\caption{\label{figure3}
The scaling plot for the Goldstone mass, $(J-J_c)L^{1/\nu}$-$Lm_G$,
 is plotted for
($+$) $N=30$,
($\times$) $32$,
($*$) $34$ and
($\Box$) $36$.
Here, the scaling parameters, $J_c=0.435$ and $\nu=0.7112$,
are taken from the existing literatures, Refs. \cite{Troyer98} and 
\cite{Hasenbusch01,Campostrini02},
respectively.
Namely, there are no adjustable parameters involved in the 
scaling analysis.
}
\end{figure}

In Fig. \ref{figure3},
we present the scaling plot for the Goldstone mass,
$(J-J_c)L^{1/\nu}$-$L m_G$,
for various $J$ and 
system sizes,
$N=30$ ($+$),
$32$ ($\times$)
$34$ ($*$)
and
$36$ ($\Box$).
Here, the scaling parameters, 
$J_c=0.435$ and 
$\nu=0.7112$, are taken from
Refs. \cite{Troyer98} and 
\cite{Hasenbusch01,Campostrini02}, respectively;
namely, there are no adjustable parameters 
involved in the scaling analysis.
From Fig. \ref{figure3},
we see that
the data collapse into a scaling curve satisfactorily 
for a considerably wide range of $J$.
Such a feature indicates that 
the simulation data already enter the scaling regime.
Encouraged by this finding,
we turn to the analysis of the spectral properties.

\subsection{\label{section2_2}
Spectral function (dynamical scalar susceptibility) $\chi_s''(\omega)$}

\begin{figure}
\includegraphics{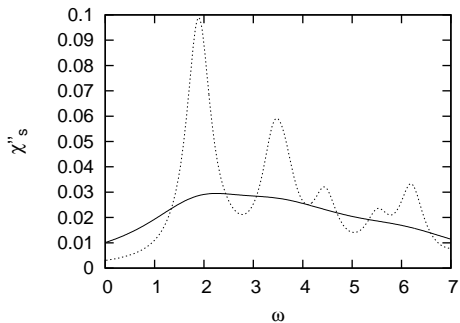}%
\caption{\label{figure4}
The spectral function
$\chi_s''(\omega)$
 (\ref{scalar_susceptibility})
is plotted for various $\omega$ with fixed 
$J=0.8$ and
$N=36$.
The $\omega$-resolution parameter $\delta$ is set to 
$\delta=1.4$ (solid)
and 
$0.3$ (dotted).
The main peak at $\omega=2.2$ (solid) corresponds to the Higgs excitation.
The main peak consists of primary ($\omega=1.9$)
and secondary ($\omega=3.5$) subpeaks;
the Higgs-excitation peak 
acquires an intrinsic width.
}
\end{figure}

In Fig. \ref{figure4},
we present the spectral function
(dynamical scalar susceptibility) \cite{Podolsky11}  
\begin{equation}
\label{scalar_susceptibility}
\chi_s''(\omega)=
-  \frac{1}{N}
{\rm Im}
   \left\langle g \left|
{\cal E}^\dagger
        \frac{1}{{\cal H}-E_g-\omega+  i \delta}
{\cal E}
   \right| g \right\rangle
    ,
\end{equation}
for various $\omega$
with fixed $J=0.8(>J_c)$ and $N=36$.
The energy-resolution parameter is set to 
$\delta=1.4$ (solid) and $0.3$ (dotted).
Here, the symbol $|g\rangle$ ($E_g$)
 denotes the ground-state vector (energy),
and the 
operator ${\cal E}$
is given by ${\cal E}= {\cal P}  {\cal H}|_{J=J_c}$
with the projection operator ${\cal P} =1-|g\rangle\langle g|$.
The spectral function $\chi_s''$ is sensitive 
(less sensitive) to the 
Higgs (Goldstone) mode,
because
the external perturbation ${\cal E}$ retains the O$(3)$ symmetry.
An advantage of the numerical diagonalization approach
is that the resolvent
$f(z)=\langle g|{\cal E}^\dagger({\cal H}-z)^{-1}{\cal E}|g\rangle$
is accessible directly
via
the continued-fraction expansion \cite{Gagliano87}.
Actually, the continued-fraction-expansion method
is essentially the same as that of the Lanczos diagonalization
algorithm (tri-diagonalization sequence), and computationally less
demanding.
The external perturbation ${\cal E}$
is
seemingly different from the conventional ones 
(implemented in the Monte Carlo simulations, for instance).
However, as far as the symmetry is concerned,
those choices are all equivalent, yielding an identical
critical behavior as to $m_H$.
Here, 
we employed the Hamiltonian itself as for ${\cal E}$,
which turned out to be less influenced by corrections to scaling.

In Fig. \ref{figure4} (solid),
we observe a Higgs-excitation peak with the mass (excitation gap),
$m_H=2.2$.
As mentioned above, the signal from the Higgs excitation
comes up, because the scalar susceptibility
$\chi_s''$ is a good probe specific to it \cite{Podolsky11};
actually, there should exist 
low-lying ($0<\omega<2.2$)
Goldstone and its continuum modes,
as illustrated in Sec. \ref{section2_1}.
Above the threshold $\omega>2.2$,
a tail background
extends.
As mentioned afterward, the present simulation was performed
so as to examine the main (Higgs) peak,
 and such high-lying spectral
 intensities 
 are beyond the scope of the present analysis.

As a reference,
we also presented
a high-resolution result
[Fig. \ref{figure4} (dotted)],
which reveals fine details of the spectral function,
namely, the series of the constituent $\delta$-function subpeaks.
The Higgs peak splits into
the primary and secondary subpeaks,
which locate at $\omega = 1.9$ and $3.5$, respectively. 
As demonstrated in the next section,
these fine structures 
(finite-size artifacts)
have to be smeared out by an adequate $\delta$
in order to attain plausible finite-size-scaling behaviors.

Last, we address a number of remarks.
First, as mentioned above,
the Higgs peak consists of two subpeaks, and 
hence,
it has an appreciable peak width.
Such feature agrees with 
the claim \cite{Rancon14}
that the Higgs peak gets broadened for the O$(N)$-symmetric model
with
large $N$.
Last, rather technically,
the continued-fraction expansion \cite{Gagliano87}
was iterated until the 
above-mentioned secondary subpeak converges.
The computational effort is comparable to
that of
the evaluation of
$|g\rangle$.

\subsection{\label{section2_3}
Finite-size-scaling analysis of $\chi_s''$}

In this section, we analyze the finite-size-scaling behavior
for $\chi_s''$ in the ordered phase, $J>J_c$.

The spectral function obeys the finite-size-scaling formula
\cite{Podolsky12}
\begin{equation}
\label{scaling}
\chi_s''(\omega)=L^{2/\nu-3} 
  f [ \omega/\Delta,(J-J_c)L^{1/\nu} ]
         ,
\end{equation}
with the critical point $J_c$,
the correlation-length critical exponent $\nu$,
a certain scaling function $f$ and the excitation gap
\begin{equation} 
\label{Delta}
\Delta(J) = m_G(2J_c-J)    ,
\end{equation}
reflected as to the critical
point $J_c$;
note that the Goldstone mass $m_G$ was considered in Sec. \ref{section2_1}.
In other words,
the Goldstone mode (in $J>J_c$)
and the fundamental energy scale $\Delta$ (in $J<J_c$)
continue adiabatically.

\begin{figure}
\includegraphics{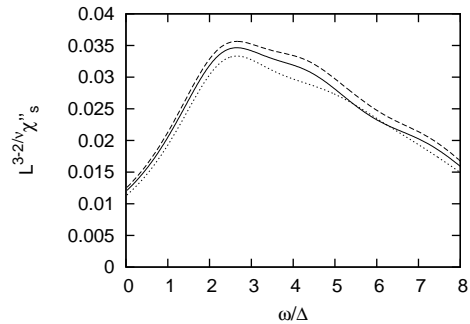}%
\caption{\label{figure5}
The scaling plot 
(\ref{scaling})
for the spectral function,
$\omega/\Delta$-$L^{3-2/\nu} \chi_s''(\omega)$,
is shown with fixed 
$(J-J_c)L^{1/\nu}=2.5$
and $\delta=1.7\Delta$
for various 
$N=32$ (dotted),
$34$ (solid) and $36$ (dashed);
here, the scaling parameters, $J_c$ and $\nu$, are the same as those
of Fig. \ref{figure3}.
The scaled Higgs peak locates at
$m_H/\Delta=2.7$.
}
\end{figure}

In Fig. \ref{figure5},
we present the scaling plot,
$\omega/ \Delta$-$L^{3-2/\nu} \chi_s'' (\omega)$,
for $N=32$ (dotted), 
$34$ (solid)
and $36$ (dashed)
with fixed $\delta=1.7\Delta$ and $(J-J_c)L^{1/\nu}=2.5$.
Here, the scaling parameters, 
$J_c=0.435$ \cite{Troyer98}
and 
$\nu=0.7112$ \cite{Campostrini02}, are
the same as those of Fig. \ref{figure3};
that is, there are no adjustable parameters in the
scaling analysis.
The scaled-spectral-function curves collapse into a scaling function $f$ satisfactorily.
From Fig. \ref{figure5},
we notice that the (properly scaled) Higgs mass 
takes a universal value $m_H/\Delta=2.7$.
The universality (stability) of $m_H/\Delta$ 
with respect to the variation of the scaling argument $(J-J_c)L^{1/\nu}$
is examined in the next section.


\subsection{\label{section2_4}
Universal character of the scaled Higgs peak}

In the above section, we investigated the universal behavior 
of $\chi_s''$
(\ref{scaling}) at a particular 
scaling argument,
$(J-J_c)L^{1/\nu}=2.5$,
and observed a scaled Higgs mass $m_H/\Delta=2.7$.
In this section, 
we vary
$(J-J_c)L^{1/\nu}$ 
in order to survey the universal
character of the Higgs peak, particularly, the scaled Higgs mass.

\begin{figure}
\includegraphics{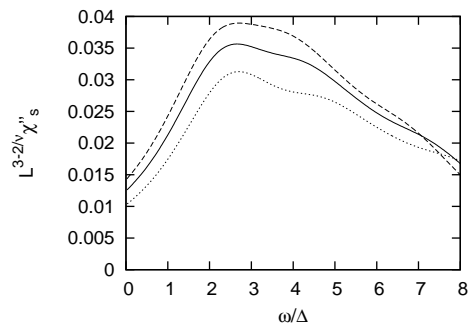}%
\caption{\label{figure6}
The scaling plot 
(\ref{scaling})
for the spectral function,
$\omega/\Delta$-$L^{3-2/\nu} \chi_s''(\omega)$,
is shown with 
fixed
$N=36$ 
and $\delta=1.7\Delta$
for various scaling arguments 
$(J-J_c)L^{1/\nu}=2$ (dotted), 
$2.5$ (solid)
and $3$ (dashed);
here, the scaling parameters, $J_c$ and $\nu$, are the same as those 
of Fig. \ref{figure3}.
The curves do not necessarily overlap,
because the scaling argument 
$(J-J_c)L^{1/\nu}$ is ranging.
The scaled Higgs-peak position 
$m_H/\Delta = 2.7$ seems to be a universal constant.
}
\end{figure}

In Fig. \ref{figure6},
we present the scaling plot, 
$\omega/\Delta$-$L^{3-2/\nu} \chi_s '' (\omega)$,
for various $(J-J_c)L^{1/\nu}=2$ (dotted), 
$2.5$ (solid) and 
$3$ (dashed) with fixed $\delta=1.7\Delta$ and $N=36$;
here,
the scaling parameters, $J_c$ and $\nu$, are the same as those of Fig. \ref{figure3}.
The data in Fig. \ref{figure6}
illustrate that 
the Higgs-peak position,
$m_H/\Delta=2.7$,
is kept invariant with respect to the variation
of $(J-J_c)L^{1/\nu}$.
On the contrary, the Higgs-peak height seems to be scattered;
note that 
according to Eq. (\ref{scaling}),
the Higgs peaks do not necessarily overlap,
because a scaling argument
$(J-J_c) L^{1/\nu}$ is no longer a constant value.
In our preliminary survey,
scanning the parameter space
$(J-J_c)L^{1/\nu}$, we observed the following tendency. 
For 
$(J-J_c)L^{1/\nu}>2$,
the scaled Higgs mass $m_H/\Delta=2.7$ is kept invariant.
In closer look, however, for 
$(J-J_c)L^{1/\nu}>3.5$,
the Higgs peak drifts to the high-energy side gradually
possibly because of the finite-size artifact (limitation of the tractable system
size).
The microscopic origin of the drift is as follows.
The spectral weight transfers from the primary subpeak 
[see Fig. \ref{figure4} (dotted)] 
to the secondary (and even ternary...) one(s)
for $(J-J_c)L^{1/\nu}>3.5$,
and the Higgs peak drifts (and gets broadened);
for exceedingly large 
$(J-J_c)L^{1/ \nu }$,
eventually, the simulation data may get out of the scaling regime.
On the one hand,
in the
$(J-J_c)L^{1/\nu}<2$ side, 
the Higgs mass acquires a significant enhancement.
This narrow regime is not physically relevant,
because
the regime shrinks
in the raw-parameter scale
[like $J-J_c(<2/L^{1/\nu})\to 0$] as $L\to\infty$.
To summarize, at least for the available system sizes $N \le 36$,
the scaling regime 
$(J-J_c) L^{1/\nu}\approx 2.5$
is optimal in the sense that
the scaled Higgs mass takes a stable 
minimal value
\begin{equation}
m_H/\Delta=2.7 .
\end{equation}

As mentioned above,
the Higgs peak consists of two subpeaks.
As a byproduct, we are able to
estimate the intrinsic peak width.
For 
$(J-J_c)L^{1/\nu} = 4.5$ and
$N=36$,
these subpeaks locate at $\omega/\Delta \approx2$ and $3.5$
with almost identical spectral weights; hence, the center locates at
$\omega/\Delta \approx 2.75$.
The distance, $1.5$, 
between these subpeaks
may be a good indicator
as to the intrinsic width of the Higgs peak,
$\delta m_H /\Delta=1.5$.
It has been claimed
\cite{Gazit13b}
that 
the Higgs peak for the O$(3)$-symmetric model
should be broadened significantly.
Our result supports this claim.

\section{\label{section3}
Summary and Discussions}

The criticality 
of the Higgs-excitation spectrum
$\chi_s''(\omega)$
[Eq. (\ref{scalar_susceptibility})]
for the bilayer Heisenberg model (\ref{Hamiltonian})
was investigated
by means of the numerical diagonalization method;
the spectral function
$\chi_s ''(\omega)$
is accessible directly via the continued-fraction expansion
\cite{Gagliano87}.
The spectral function appears to
obey the scaling formula (\ref{scaling}) satisfactorily,
indicating that the simulation data already enter the scaling regime.
As a result,
we estimated the 
scaled Higgs mass 
$m_H/\Delta=2.7$ with the
peak width $\delta m_H /\Delta=1.5$.
So far, with the (quantum) Monte Carlo method,
the scaled Higgs mass has been estimated as 
$m_H/\Delta = 2.2(3)$ \cite{Gazit13a}
and 
$2.6(4)$ \cite{Lohofer15}.
According to the normalization-group analysis,
the scaled Higgs mass was estimated as 
$m_H /\Delta = 2.7$ \cite{Rose15}
and 
$1.64$ \cite{Katan15}.
Our result agrees with these preceding estimates
\cite{Gazit13a,Lohofer15,Rose15};
the error margin of our estimate should be bounded by half
a peak width, $\approx 0.75$.

The Ginzburg-Landau theory
(based on the wine-bottle-bottom potential)
yields the critical amplitude ratio
$m_H/\Delta=\sqrt{2}$.
Clearly, the Ginzburg-landau theory fails in describing
the spectral property for the $d=3$ O$(3)$ universality class.
In other words,
such a spectral property reflects a character of each universality class
rather sensitively.
As a matter of fact,
as for the ``deconfined critical'' phenomenon \cite{Huh13},
an exotic spectral property was predicted.
A consideration toward this direction is left for the future study.

\section*{Acknowledgment}
This work was supported by a Grant-in-Aid
for Scientific Research (C)
from Japan Society for the Promotion of Science
(Grant No. 25400402).

\appendix

\section{\label{appendix}Numerical algorithm: Screw-boundary condition \cite{Novotny90}}

In this Appendix, 
we explain the simulation algorithm to 
diagonalize the Hamiltonian matrix
for the 
bilayer Heisenberg model (\ref{Hamiltonian}).
We implemented the screw-boundary condition
\cite{Novotny90}, with which one is able to treat 
a variety of system sizes $N=30,32,\dots$ ($N$: the number of constituent spins)
systematically.
According to Ref. \cite{Novotny90},
an alignment of spins $\sigma_i$ ($i=1,2,\dots,M$)
with both nearest- and $\sqrt{M}$th-neighbor interactions
is equivalent to
a two-dimensional cluster under the screw-boundary condition;
here, the periodical boundary condition as to the spin alignment,
namely, $\sigma_{M+i}=\sigma_i$, is imposed.
Based on this idea, we express the Hamiltonian matrix 
\begin{eqnarray}
{\cal H}
&=& -J \sum_{a=1}^{2} \sum_{i=1}^{N/2} 
     (P^{-\sqrt{N/2}}{\bf S}_{ai}P^{\sqrt{N/2}})\cdot {\bf S}_{ai}
 -J \sum_{a=1}^{2} \sum_{i=1}^{N/2} {\bf S}_{a,i+1}\cdot{\bf S}_{ai} \nonumber \\
\label{Hamiltonian_Novotny}
 &&
 +J' \sum_{i=1}^{N/2} {\bf S}_{1i}\cdot{\bf S}_{2i} 
    ,
\end{eqnarray}
with the translation operator (by one lattice spacing) $P$ \cite{Novotny90};
namely, a relation $P^{-\delta} {\bf S}_{ai} P^{\delta}={\bf S}_{a,i+\delta}$ holds.
We diagonalized the above Hamiltonian matrix
(\ref{Hamiltonian_Novotny}) with the Lanczos
method so as to evaluate the ground-state vector (energy) $|g\rangle$ ($E_g$).
The above expression (\ref{Hamiltonian_Novotny}) is mathematically closed.
However, as for an efficient simulation,
a formula
(11) of Ref. 
\cite{Nishiyama08}
may be of use in order to cope with the operation
$P^{\pm \sqrt{N/2}}$.

%
%
%
%

\end{document}